\documentclass[english,aps,prb,superscriptaddress,reprint]{revtex4-2}
\usepackage[T1]{fontenc}
\usepackage[latin9]{inputenc}
\usepackage{babel}
\usepackage{graphicx}
\usepackage{bm}
\usepackage{amsmath}
\usepackage{amssymb}
\usepackage{amsfonts}
\usepackage{dcolumn}
\usepackage{bbold}
\usepackage{tikz}
\usepackage{xfp}
\usepackage{verbatim}

\usepackage[margin=15mm]{geometry}
\colorlet{linkequation}{blue}
\usepackage[colorlinks=true,linkcolor=blue,citecolor=blue,urlcolor=blue]{hyperref}
\usetikzlibrary{shadings}

\usepackage{color}

\begin{document}

\begin{abstract}
We investigate spin pumping current injected by the nutation resonances of a ferromagnet or an antiferromagnet into an adjacent metal. Comparing the dc spin pumping current between the normal precession and nutation resonances,  we find that the ratio of spin pumping current at the nutation resonance to the precession resonance is more pronounced in antiferromagnets. We further show that the spin pumping current injected by the nutation resonance is opposite in sign as compared to the normal precession mode. This could offer a useful experimental signature for identifying such nutation resonances. Analyzing the nature of the nutational eigenmodes, we show that the sign change in spin current is rooted in a reversal of the precession sense for the nutation mode(s). Furthermore, the nutational modes in antiferromagnets are found to be dominated by precession of one of the two sublattices only.

\end{abstract}

\title{Spin pumping at {terahertz} nutation resonances}

\author{Ritwik Mondal}
\email[]{mondal@fzu.cz}

\affiliation{Department of Spintronics and Nanoelectronics, Institute of Physics of the Czech Academy of Sciences, Cukrovarnick\'a 10, CZ - 162 00 Praha 6, Czech Republic}

\affiliation{Department of Physics and Astronomy, Uppsala University, Box 516, Uppsala, SE-75120, Sweden}

\author{Akashdeep Kamra}
\email[]{akashdeep.kamra@uam.es}
\affiliation{Condensed Matter Physics Center (IFIMAC) and Departamento de F\'{i}sica Te\'{o}rica de la Materia Condensada, Universidad Aut\'{o}noma de Madrid, E-28049 Madrid, Spain}

\maketitle
\date{\today}

{ \section{ Introduction}}
The ultrafast manipulation of spins at the terahertz frequencies has paramount scientific and technological interest. Most spin dynamics experiments have been explained so far by the traditional Landau-Lifshitz-Gilbert (LLG) phenomenology \cite{Bigot1996,kampfrath13,Walowski2016,John2017,ostler12,Wienholdt2012,Wienholdt2012PRL},  which describes a precessional motion of spins around an effective field and an energy dissipation via a viscous damping term~\cite{landau35,gilbert04}. However, at ultrafast timescales, the LLG equation has been found to be insufficient~ \cite{Blank2021THz,Mondal2019PRB}. 

To address this shortcoming, the LLG phenomenology has been extended to account for magnetization dynamics in the inertial regime~\cite{Ciornei2011,Wegrowe2012}. Essentially, the inclusion of magnetic inertia leads to a spin nutation at the ultrashort timescales and can be incorporated via a double time-derivative of the magnetization i.e., $\bm{M}\times \ddot{ \bm{M}}$~\cite{Suhl1998,Bottcher2012}. The inertial LLG (ILLG) equation of motion for two-sublattice systems has the form
\begin{align}
\dot{\bm{M}}_i & = -\gamma_i \left(\bm{M}_i\times \bm{H}^{\rm eff}_{i}\right) + \frac{\bm{M}_i}{M_{i0}}\times\left[\alpha_i \dot{\bm{M}}_i + \eta_i \ddot{ \bm{M}}_i\right]\,,
\end{align}   
with the gyromagnetic ratio $\gamma$, the effective field $\bm{H}^{\rm eff}$, ground state magnetic moment $M_0$, the Gilbert damping parameter $\alpha$ and the inertial relaxation time $\eta$. The index ``$i$'' denotes the sublattice. In general, Gilbert damping in a two-sublattice and the inertial relaxation time are tensors~\cite{Mondal2016,Mondal2018PRB,Kamra2018,Nieves2015,Yuan2019}. Here, we treat them as scalars for simplicity and in an attempt to capture the leading order effects. The emergence of spin nutation has been attributed to several proposed mechanisms \cite{Fahnle2011JPCM,Fahnle2011,Bhattacharjee2012,Bajpai2019PRB,Mondal2017Nutation,Mondal2018JPCM,Giordano2020,Titov2021}.
The characteristic timescales of the nutation $\eta$ have been predicted to {be in} a range of $1-100$ fs~\cite{Li2015,Makhfudz2020}. Experimentally inertial relaxation time $\eta$ is found to be about hundreds fs in two sublattice ferromagnets~\cite{Neeraj2021}. 

The spin nutation additionally introduces a second resonance in the ferromagnetic resonance (FMR) spectrum, however, at a higher frequency in THz range. Such a resonance is called ferromagnetic nutation resonance (FMNR) \cite{Olive2015,cherkasskii2020nutation}. Moreover, the precession resonance frequencies are decreased due to the spin nutation \cite{Olive2015, Makhfudz2020,MondalJPCM2021}. In a more recent experiment, several higher-order nutation resonances have been observed \cite{unikandanunni2021inertial}. With rapid recent progress, new pathways for the role of nutation in practical devices have already started to emerge~\cite{Rahman2021}.

Despite several signatures in ultrafast spin dynamics experiments attributed to nutation modes \cite{Li2015,Neeraj2021,unikandanunni2021inertial}, further smoking-gun validations are needed. Complementary to a time-resolved tracking of the magnetization dynamics, the spin pumping current~\cite{Tserkovnyak2002,Saitoh2006} injected by the latter into an adjacent metal has emerged as a powerful probe. It has already been employed in investigating a broad range of phenomena from spin Seebeck effect~\cite{Uchida2008,Bauer2012,Barker2016,Seifert2018,Hirobe2017,Weiler2013} to antiferromagnetic resonance~\cite{Cheng2014,Vaidya2020,Li2020}, and over a broad range of time scales. With the anticipation that spin pumping current driven by the nutation mode might be very different from the conventional resonance modes, we theoretically investigate this hypothesis in both ferromagnets (FM) and antiferromagnets (AFM) here finding valuable results and insights.

In this article, we theoretically investigate the dc spin pumping current injected by a magnet driven by an oscillating magnetic field into an adjacent metal. We consider both ferro- and antiferromagnets, and focus on their nutation modes. We find that the spin current at the nutation resonance is negative, while the spin current at the precession resonance is positive. Our results also show that the spin current at the nutation resonance increases with increasing inertial relaxation time $\eta$ in FM and AFM. The computed ratio of spin currents at the nutation to the precession resonance show that the nutation spin current is more pronounced in AFM. We delineate the nutation eigenmodes finding the ferromagnetic nutation to entail magnetization precession in the opposite sense, compared to the normal precession mode. In AFM, the nutation modes are characterized by a larger precession cone for one of the two sublattice magnetizations, thereby departing from the nearly collinear dynamics in the conventional antiferromagnetic resonance. These features may enable a clear distinction of the nutation modes in experiments.

{ \section{ Spin pumping in ferromagnets}}
We consider a FM with a single sublattice as $\bm{M} = M_0\hat{\bm{z}}$ at the ground state, that is under the influence of an external Zeeman field $\bm{H} = H_0\hat{\bm{z}}$.
Such a FM can be described by the following free energy $\mathcal{F}(\bm{M}) = -H_0M_z - KM_z^2/M_0^2 $, where $K$ is the uniaxial anisotropy energy and $M_0$ is the ground state magnetic moment. The effective field that enters into the ILLG equation can thus be calculated using $\bm{H}^{\rm eff} = -\partial \mathcal{F}/\partial \bm{M}$. 

When a small oscillating transversal field $\bm{h}(t) = h_x(t) \hat{\bm{x}} + h_y(t)\hat{\bm{y}}$ is applied, the small magnetization oscillations are induced such that  time-dependent magnetization is $\bm{M}(t) =  m_x(t)\bm{\hat{x}} + m_y(t)\bm{\hat{y}} + M_0\hat{\bm{z}}$. Within the linear response theory in the circular basis described by $h_{\pm} = h_x \pm { i} h_y = he^{\pm{ i} \omega t}$ and $m_{\pm} = m_x \pm { i} m_y = me^{\pm{ i}\omega t}$, the calculated susceptibility expression is \cite{Mondal2020nutation}
\begin{align}
    m_\pm & = \frac{\gamma M_0}{\Omega_0 -\eta \omega^2 - \omega \pm { i} \omega\alpha }h_\pm = \chi_{\pm}h_\pm \,,
    \label{ferro_susceptibility}
\end{align}
where $\Omega_0 = \frac{\gamma}{M_0} \left[H_0M_0 + 2K\right]$. The poles of the susceptibility determine the resonance frequencies. Without the nutation term $\eta$, only a FMR frequency is obtained. However, the spin nutation additionally introduces a second resonance FMNR frequency. These two frequencies are  \cite{Mondal2020nutation} 
\begin{align}
    \omega_{\rm FMR} & = \frac{\sqrt{1 + 4\eta \Omega_0} - 1}{2\eta}\,,\\
    \omega_{\rm FMNR} & = -\frac{\sqrt{1 + 4\eta \Omega_0} + 1}{2\eta}\,.
\end{align} 
The negative frequency in the nutation resonance dictates the fact that the nutation resonance has opposite handedness of rotation compared to the FMR~\cite{Kikuchi,Mondal2020nutation}. This has been further corroborated by an analysis of the nutation eigenmode presented in the {Appendix \ref{appendixa}}. 

\begin{figure}[tbh!]
    \centering
    \includegraphics[scale = 0.43]{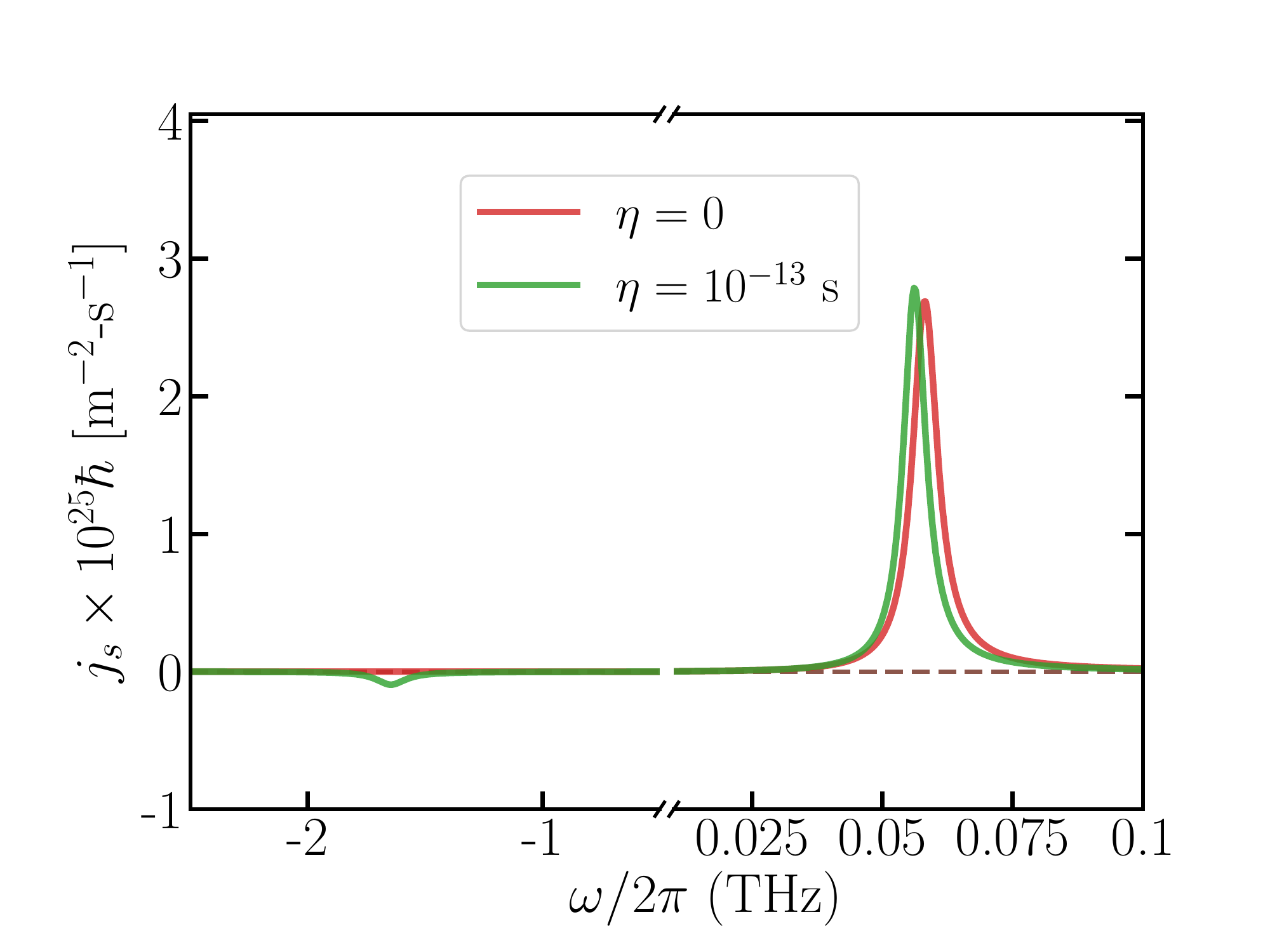}
    \caption{The calculated spin pumping dc current for inertial relaxation times $\eta = 0$ s and  $\eta = 10^{-13}$ s. The used parameters are $M_0 = 2 \mu_B$, $K = 10^{-23} $ J, $\gamma = 1.76\times 10^{11}$ T$^{-1}$-s$^{-1}$, $\alpha = 0.05$, $H_0 = 1$ T, $\vert h \vert = 10^{-3}$ T, $g_r^{\uparrow\downarrow} = 10^{19}$ m$^{-2}$. }
    \label{fig1}
\end{figure}

The dc component of a generated spin current density can be expressed as~\cite{Ando2011,Tserkovnyak2002}
\begin{align}
    j_s & = \frac{\omega}{2\pi} \int_0^{2\pi/\omega} \frac{\hbar}{4\pi} g_r^{\uparrow\downarrow} \frac{1}{M_0^2} \left[\bm{M}(t)\times {\dot{\bm{M}}(t)}\right]_z \,dt\,.
    \label{current}
\end{align}
We calculate the spin current in the circular basis and $\left[\bm{M}(t)\times {\dot{\bm{M}}(t)}\right]_z = \frac{ i}{2}\left[m_+\dot{m}_--m_-\dot{m}_+\right].  $ 
Using Eq.\ (\ref{ferro_susceptibility}), the calculated spin current has the expression
\begin{align}
    j_s & = \frac{\hbar}{4\pi} g_r^{\uparrow\downarrow}\left[\frac{ \omega \gamma^2}{(\Omega_0 - \eta\omega^2  - \omega)^2 + \alpha^2 \omega^2 } \vert h\vert^2\right]\,.
    \label{js_ferromagnet}
\end{align}
{We compute such spin currents with inertial relaxation time $\eta = 10^{-13}$ s in Fig. \ref{fig1}. 
}
We denote the calculated spin current at FMR frequencies as $j_s^{\rm FMR}$ and at nutation frequencies as $j_s^{\rm FMNR}$.
Note that the spin current at the FMR frequencies has positive sign, however, the calculated spin current at the nutation resonance frequencies  has the opposite sign. The reason is that while the precession mode rotates anticlockwise, the nutation mode rotates clockwise. 
\begin{figure}[tbh!]
    \centering
    \includegraphics[scale = 0.25]{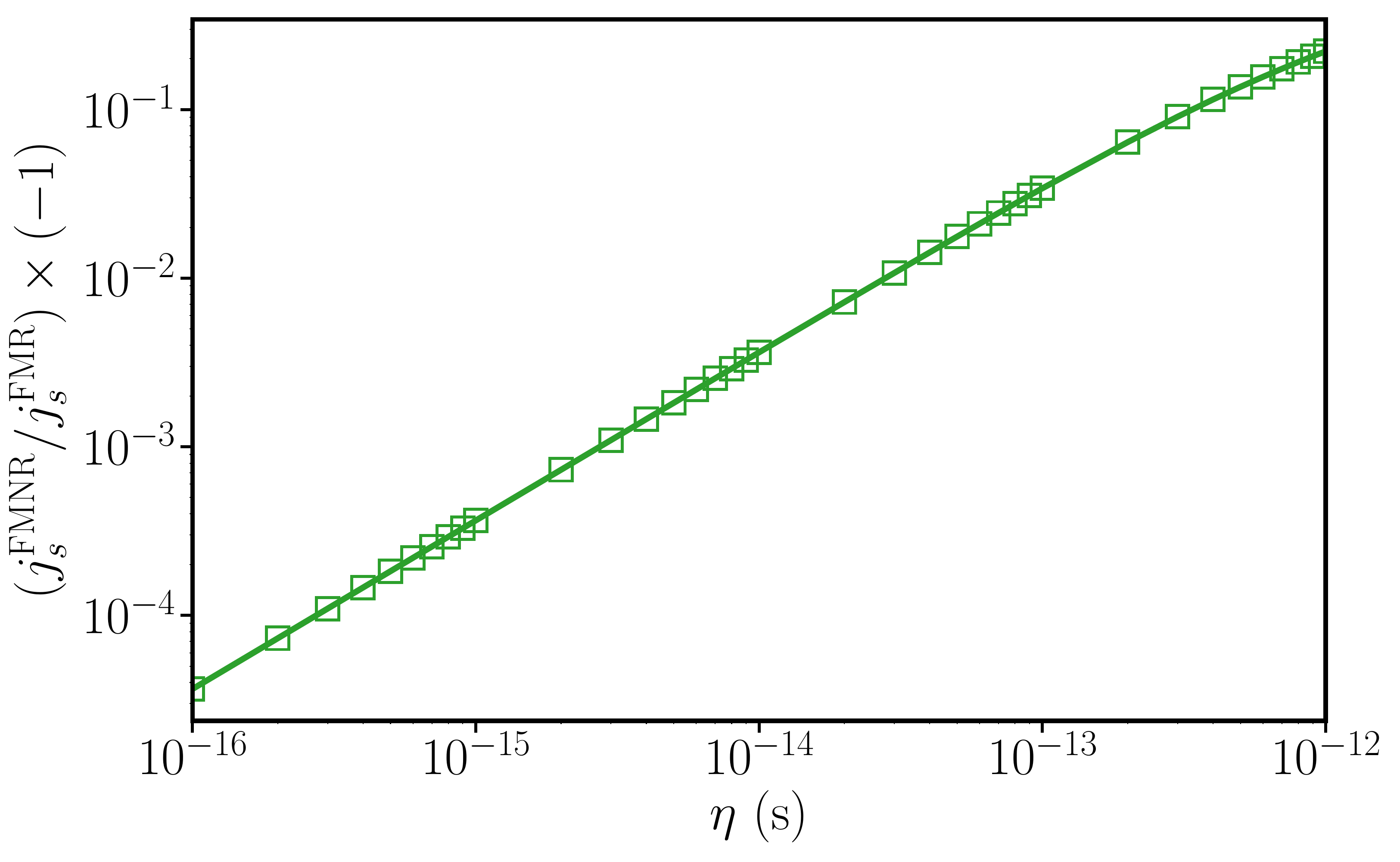}
    \caption{The ratio of spin current for ferromagnets at the nutation resonance to the precession resonance vs inertial relaxation time $\eta$. {The used parameters are $M_0 = 2\mu_B$, $\gamma = 1.76\times 10^{11}$ T$^{-1}$-s$^{-1}$,  $\alpha = 0.05$, $K = 10^{-23}$ J and $H_0 = 0$ T.}}
    \label{fig2}
\end{figure}

Nonetheless, we calculate the ratio of spin currents calculated at several inertial relaxation times $\eta$ in Fig.\ \ref{fig2}. {Note that the ratio of spin currents is independent of several parameters including spin mixing conductance $g^{\uparrow\downarrow}_r$, small field $h$ etc.} We emphasize  that the spin current at the nutation resonance increases linearly with $\eta$, while the current at the precession resonance stays almost constant. Such an observation can easily be understood from Eq.\ (\ref{js_ferromagnet}). The FMR frequency lies in the GHz regime and its shift due to nutation is small in ferromagnets. Therefore, the calculated spin current at the FMR is roughly proportional to $\omega_{\rm FMR}^{-1} \approx 1/\Omega_0 + \eta$. However, the dominant spin current contribution at the FMNR is $\omega_{\rm FMNR}^{-1} \approx -\eta$. Therefore, at smaller $\eta$, the ratio of spin currents is linear in $\eta$, however, the ratio deviates from linearity at larger $\eta$.

{ \section{ Spin pumping in antiferromagnets}}
We consider an AFM having two sublattices namely $A$ and $B$. The ground state of such AFM is $
\bm{M}_{A} = M_{A0}\hat{\bm{z}}$ and $
\bm{M}_{B} = -M_{B0}\hat{\bm{z}}$ under an influence of external applied Zeeman field $\bm{H} = H_0\hat{\bm{z}}$. 
The free energy of the system \cite{Mondal2020nutation,Kamra2018}
\begin{align}
    & \mathcal{F}\left( \bm{M}_A, \bm{M}_B\right)  = - H_0 \left(M_{Az} + M_{Bz}\right) \nonumber\\
    & - \frac{K_A}{M_{A0}^2}M_{Az}^2 - \frac{K_B}{M_{B0}^2}M_{Bz}^2 +\frac{J}{M_{A0}M_{B0}} \bm{M}_A \cdot \bm{M}_B \,,
\end{align}
containing the Zeeman field, uniaxial anisotropy energies for individual sublattices in terms of $K_A$ and $K_B$, and the inter-sublattice magnetic exchange energy $J$. The effective field in the ILLG equation can be calculated using the following definition: $\bm{H}^{\rm eff}_A = -\partial \mathcal{F}/\partial \bm{M}_A$  and $\bm{H}^{\rm eff}_B = -\partial \mathcal{F}/\partial \bm{M}_B$. 
Similar to our consideration of FM, we calculate the AFM susceptibility assuming $\bm{h}_A = \bm{h}_B$ and $\bm{M}_A(t) = M_{A0}\hat{\bm{z}} + \bm{m}_A(t)$ and $\bm{M}_B(t) = -M_{B0}\hat{\bm{z}} + \bm{m}_B(t)$. 
In the circular basis, the inverse susceptibility expression for AFM is obtained as \cite{Mondal2020nutation}
\begin{widetext}
\begin{align}
    \begin{pmatrix}h_{A\pm} \\ h_{B\pm}\end{pmatrix}=  \begin{pmatrix}\dfrac{1}{ \gamma_A M_{A0}}\left(\Omega_A \pm {i}\omega\alpha_{A} -\eta_{A}\omega^2-\omega  \right) &  \dfrac{J}{M_{A0}M_{B0}}\\
      \dfrac{J}{M_{A0}M_{B0}} & 
     \dfrac{1}{ \gamma_B M_{B0}}\left(\Omega_B \pm {i}\omega\alpha_{B} -\eta_{B}\omega^2 +\omega  \right)  
    \end{pmatrix}\begin{pmatrix}m_{A\pm} \\ m_{B\pm}\end{pmatrix}\,,\label{suscAFM}
\end{align}
\end{widetext}
with the following definitions $\Omega_A  =  \frac{\gamma_A}{M_{A0}}( J  + 2K_A +  H_0 M_{A0})$, 
$\Omega_B  = \frac{\gamma_B}{M_{B0}}( J + 2K_B - H_0 M_{B0})
$. For an AFM, we consider, $\gamma_A = \gamma_B = \gamma$, $M_{A0} = M_{B0} = M_0$, $K_A = K_B = K$, $\eta_A = \eta_B = \eta$. With the assumption that $J \gg H_0M_0$ and $J\gg K$, we have $\Omega_A \approx \Omega_B \approx \gamma J / M_0$. Therefore, the obtained approximate frequencies of the antiferromagnetic precession resonance (AFMR) and antiferromagnetic nutation resonance (AFMNR) are \cite{Mondal2020nutation,Mondal2021PRBNutation}
\begin{align}
    \omega_{\rm AFMR} & \approx \pm \frac{\gamma}{M_0} \sqrt{\dfrac{{4JK}}{1+ \dfrac{2\eta\gamma J}{M_0}}}\,,\\
    \omega_{\rm AFMNR} & \approx \pm \frac{\sqrt{1 +\dfrac{2\eta\gamma J}{M_0}}}{\eta}\,.
\end{align}
Note that unlike FM, the AFM has two sublattices meaning there are two precession resonance frequencies and corresponding two nutation frequencies. { However, we also mention that in a two sublattice FM  there are two of each precession and nutation resonance frequencies e.g., CoFeB \cite{Neeraj2021,MondalJPCM2021}.}

 \begin{figure}[tbh!]
    \centering
    \includegraphics[scale = 0.32]{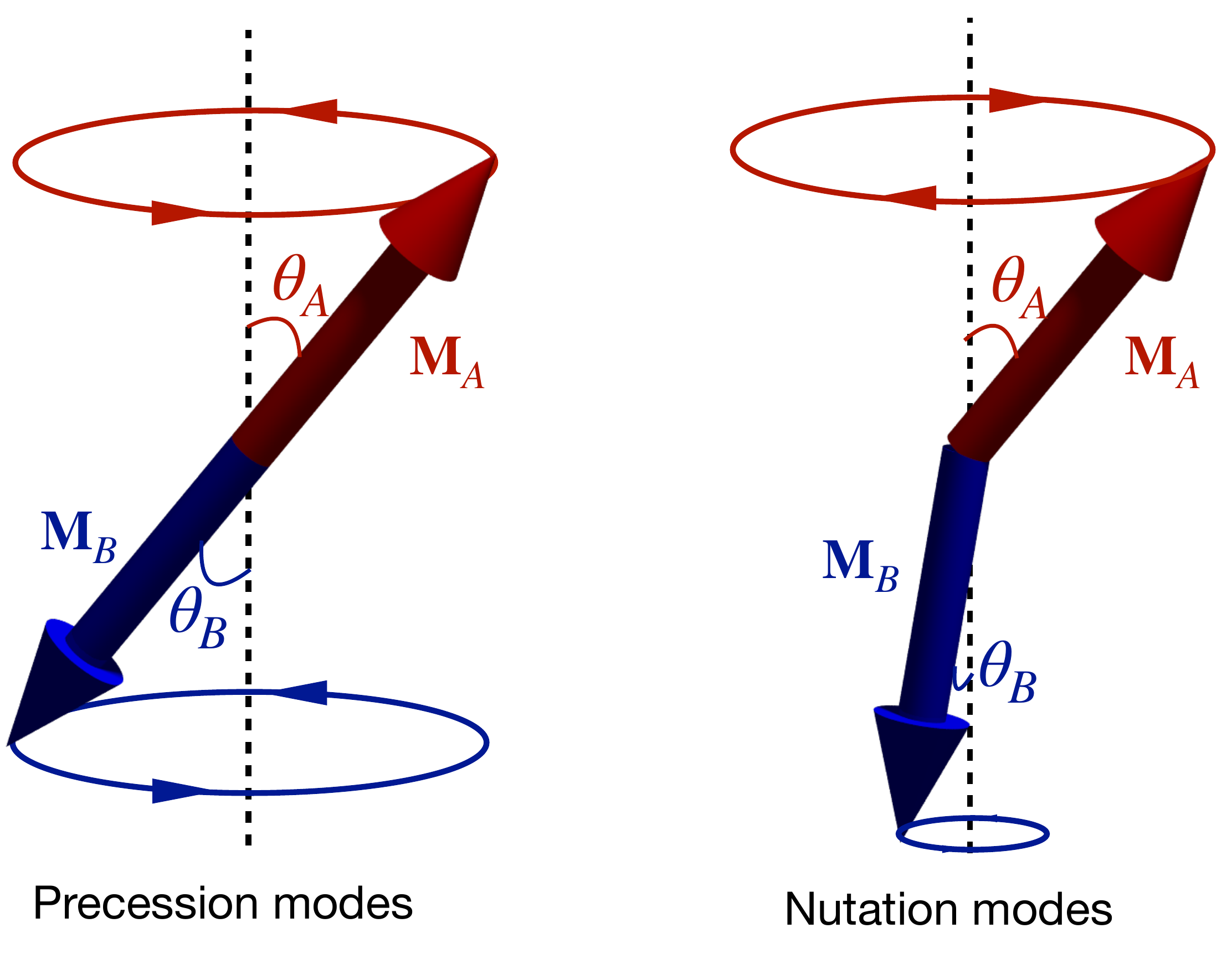}
    \caption{The precession and nutation modes in antiferromagnets. The precession modes behave as normal antiferromagnets, but for the nutation mode, the spins are not antiparallel giving rise to faster dynamics in one sublattice, while the dynamics is not experienced by the other sublattice.}
    \label{AFM_Sketch}
\end{figure} 

First we examine and compare the conventional and nutational eigenmodes in an AFM. To this end, we set $h_{A\pm} = h_{B\pm} = 0$ and $\alpha = 0$ in Eq.\ (\ref{suscAFM}) and obtain two following equation for $m_{A+}$ and $m_{B+}$:
\begin{align}
    \left(\Omega_A - \eta\omega^2 - \omega\right) m_{A+} + \frac{\gamma J}{M_{0}} m_{B+} & = 0\,,\\
    \left(\Omega_B - \eta\omega^2 +\omega\right)m_{B+} +  \frac{\gamma J}{M_{0}} m_{A+} & = 0\,.
\end{align}
Two of such similar equation can also be obtained for $m_{A-}$ and $m_{B-}$.
At AFMR, we drop the nutation term and  the ratio between $m_{A+}$ and $m_{B+}$  is obtained as
\begin{align}
    \frac{m_{A+}}{m_{B+}} & = -\frac{\dfrac{\gamma J}{M_0}}{\Omega_A - \omega_{\rm AFMR}} \approx -\frac{1}{1- \frac{\omega_{\rm AFMR} M_0}{\gamma J}}\,.
\end{align}
 One can easily calculate that $\frac{\omega_{\rm AFMR}M_0}{\gamma J} \ll 1$ and can be neglected. Therefore, we obtain $m_{A+} \approx - m_{B+}$ implying that the two sublattice magnetizations remain nearly antiparallel subtending almost equal cone angles $\theta_A \approx \theta_B$ in AFMR, as depicted in Fig.~\ref{AFM_Sketch} (left figure).

For nutation mode, we similarly obtain the ratio with the leading term in AFMNR frequency as $\omega_{\rm AFMNR} \approx 1/\eta$ as
\begin{align}
    \frac{m_{A+}}{m_{B+}} & = -\frac{\dfrac{\gamma J}{M_0}}{\Omega_A - \eta\omega^2_{\rm AFMNR}- \omega_{\rm AFMNR}} = \frac{1}{ \frac{2M_0}{\eta \gamma J} - 1}\,.
\end{align}
Employing $\frac{M_0}{\eta \gamma J}\gg 1$, we find $m_{A+} \ll m_{B+}$. This means at the AFMNR, the two sublattices do not align exactly antiparallel, giving rise to large cone angle in one sublattice, while the other sublattice precesses with a much smaller cone angle. These characteristics of the nutation eigenmodes have been shown in Fig.\ \ref{AFM_Sketch} (right figure). 
 
\begin{figure*}[tbh!]
\centering
\includegraphics[scale = 2]{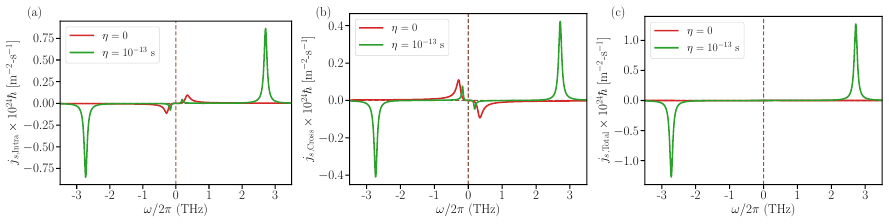}
\caption{The spin current contributions for (a) intra-sublattice terms: $\bm{M}_A(t)\times \dot{\bm{M}}_{A}(t) + \bm{M}_B(t)\times \dot{\bm{M}}_{B}(t)$ (b) cross-sublattice terms: $\bm{M}_A(t)\times \dot{\bm{M}}_{B}(t) + \bm{M}_B(t)\times \dot{\bm{M}}_{A}(t)$ and (c) total with both intra and cross-sublattice terms calculated  without and with the inertial relaxation time i.e., $\eta= 0$ and $\eta = 10^{-13}$ s. {The used parameters are: $M_{A0} = M_{B0} = M_0 = 2 \mu_B$, $\gamma_A = \gamma_B = \gamma = 1.76\times10^{11}$ T$^{-1}$-s$^{-1}$, $\hbar = 1.05\times 10^{-34}$ m$^2$-kg-s$^{-1}$, $g_r^{\uparrow \downarrow} = 10^{-19}$ m$^{-2}$, $H_0 = 0$ T, $h_A = h_B = 10^{-3}$ T, $K_A = K_B = K = 10^{-23}$ J, $J = 10^{-21}$ J, $\alpha_A = \alpha_B = \alpha = 0.05$ and $\eta_A = \eta_B = \eta$.}}
\label{Antiferro_spin_current}
\end{figure*}

Next, we calculate the spin pumping current following Refs.~\cite{Kamra2017,Liu2017}. To this end, we consider the effects of intra-sublattice and also cross-sublattice terms in the spin pumping current. The spin mixing conductance for an AFM interfaces with a metal is a $2\times2$ tensor and depends sensitively on the interface~\cite{Kamra2017,Troncoso2021}. A disordered interface (expected to be common) effectively behaves as an uncompensated interface~\cite{Kamra2017,Troncoso2021}. Here, we report the contributions due to the intra-sublattice and cross-sublattice spin pumping separately, keeping in mind that the final result is a weighted sum of the two, where the weight depends on the interface~\cite{Kamra2017,Troncoso2021}. The intra-sublattice contributions can be calculated as: $\bm{M}_A(t)\times \dot{\bm{M}}_{A}(t) + \bm{M}_B(t)\times \dot{\bm{M}}_{B}(t) = \omega \left(m_{A+}m_{A-} + m_{B+}m_{B-}\right)$ and the cross-sublattice contributions as: $\bm{M}_A(t)\times \dot{\bm{M}}_{B}(t) + \bm{M}_B(t)\times \dot{\bm{M}}_{A}(t) = \omega \left(m_{A+}m_{B-} + m_{B+}m_{A-}\right)$. {The computed spin current due to intra and cross-sublattice contributions become
\begin{align}
    j_{ s,\rm Intra} 
    & = \frac{\hbar}{4\pi} g_r^{\uparrow\downarrow}\times \omega\left(\frac{m_{A-}{m}_{A+}}{M_{A0}^2} +\frac{m_{B-}{m}_{B+}}{M_{B0}^2} \right)\\
     j_{ s,\rm Cross} 
    & = \frac{\hbar}{4\pi} g_r^{\uparrow\downarrow}\times \omega\left(\frac{m_{A-}{m}_{B+}}{M_{A0}M_{B0}} +\frac{m_{B-}{m}_{A+}}{M_{B0}M_{A0}} \right)
\end{align}
The total spin current is calculated from both the intra and cross-sublattice contributions as 
\begin{align}
    j_{s,\rm Total} & = j_{s,\rm Intra} + j_{s,\rm Cross} 
\end{align}
 } 
The computed spin currents are shown for intra, cross, and total sublattice terms at the inertial relaxation time $\eta = 100$ fs in Fig.\ \ref{Antiferro_spin_current}. Without the application of a static magnetic field, the two antiferromagnetic precession resonance modes have exactly the same frequency, however, opposite in sign. Therefore, the spin currents appear at the resonance frequencies.  Due to the nutation resonance, additional spin current contributions can be observed at the higher THz nutation frequencies. Following the results of FM case, the sign of nutation spin currents will be opposite to the corresponding spin current at the precession resonance mode, for intra-sublattice contributions. 

As observed, the magnitude of spin currents at the nutation resonances is small compared to the spin currents at the precession resonance at the lower $\eta$ e.g., 1 fs. However, it increases rapidly for higher $\eta$ and surpasses the spin currents at the precession resonance already at $\eta \sim 10 $ fs. Such observation in AFMs is in contrast to the FM, where the nutation spin current is smaller even at $\eta = 1$ ps. Note that the antiferromagnetic precession resonance is suppressed and thus the spin current at the precession resonance is smaller compared to that of the ferromagnetic resonance.

For intra-sublattice contributions, the nutation spin current has opposite sign to the precession spin current, consistent with our findings for ferromagnets [see Fig.\ \ref{Antiferro_spin_current}(a)]. However, for the cross-sublattice contributions, the precession spin current changes sign, while the nutation spin current does not change sign compared to the intra-sublattice contributions [see Fig.\ \ref{Antiferro_spin_current}(b)]. This is due to the fact that the two sublattices in the antiferromagnetic precession mode remain almost antiparallel to each other. On the other hand, for the nutation modes in AFM, the two sublattices do not align antiparallel to each other and hence the nutation spin current do not change sign for intra and cross-sublattice contribution. Due to such properties, the total spin currents at the precession resonance almost cancel with each other, however, the total nutation spin currents add up [see Fig.\ \ref{Antiferro_spin_current}(c)], assuming equal intra- and cross-sublattice spin mixing conductances.
\begin{figure}[tbh!]
    \centering
    \includegraphics[scale = 0.27]{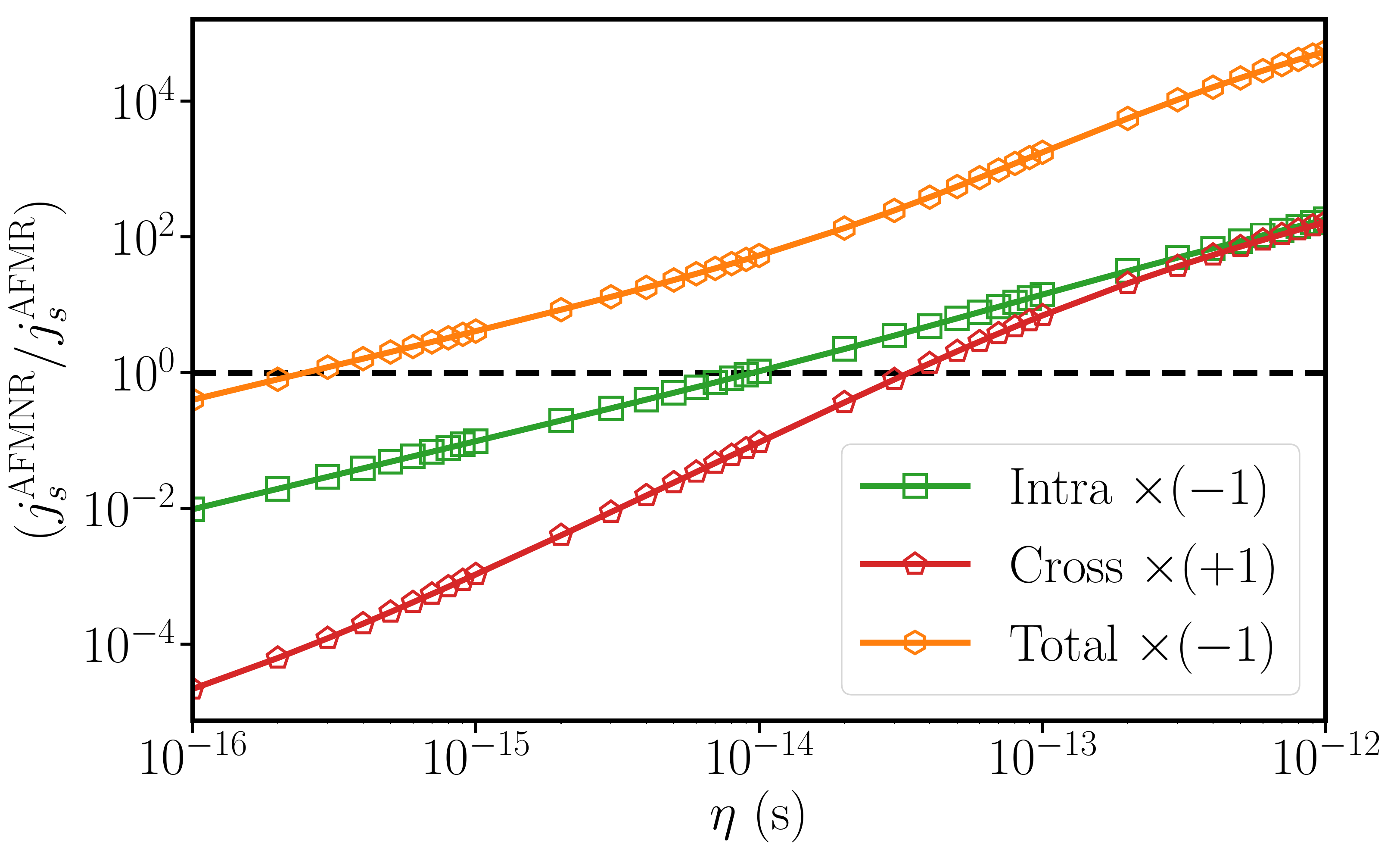}
    \caption{The ratio of spin current for AFMs at the nutation resonance to the precession resonance vs inertial relaxation time. {The used parameters are: $M_{A0} = M_{B0} = M_0 = 2 \mu_B$, $\gamma_A = \gamma_B = \gamma = 1.76\times10^{11}$ T$^{-1}$-s$^{-1}$, $H_0 = 0$ T, $K_A = K_B = K = 10^{-23}$ J, $J = 10^{-21}$ J, $\alpha_A = \alpha_B = \alpha = 0.05$ and $\eta_A = \eta_B = \eta$.}
    }
    \label{Ratio_AFM}
\end{figure}

To understand the enhancement of spin currents at the  nutation resonance, we compute the ratio of spin currents at the nutation resoance to the precession resonance in Fig.\ \ref{Ratio_AFM}. By doing so, the results are independent of several parameters used in obtaining Fig.\ \ref{Antiferro_spin_current}. For example, the ratio of spin currents is independent of spin mixing conductance $g_r^{\uparrow\downarrow}$, small field $h$ etc.  Compared to the FM spin current in Fig.\ \ref{fig1}, the such ratio already crosses unity within $\eta = 10$ fs for AFM.  As explained earlier, due to the cancellation of the spin current at the precession resonance, the ratio for total contribution increases rapidly. In fact, the total spin current at the nutation resonance is already higher even below $\eta = 1$ fs.  

{ \section{Conclusions}}
To conclude, we present a theoretical investigation of spin pumping with the magnetic inertial dynamics for one and two-sublattice systems. The magnetic inertial dynamics additionally introduces a spin pumping current at the THz nutation resonance frequencies. However, due to the opposite sense of rotation in precession and nutation modes, the spin pumping current has an opposite sign at the nutation resonance compared to the one at precession resonance. Such scenario remains the same for intra-sublattice spin current in AFM, while the cross-sublattice spin current only changes sign at the precession resonance. Thus the spin current ratio is negative and positive for AFM intra and cross-sublattice terms, respectively. { While  the two sublattices remain almost antiparallel in AFM at the precessional resonance, the antiparallel alignment is broken at the nutational resonance resulting a large cone angle in one of the sublattices. } 
Our obtained results should motivate to the experimental search of magnetic inertial dynamics in magnets via detection of spin pumping currents.      

{ \section*{ Acknowledgments}}
We acknowledge  the Swedish Research Council (VR Grant No. 2019-06313) and the Spanish Ministry for Science and Innovation
- AEI Grant CEX2018-000805-M (through the ``Maria de Maeztu'' Programme for Units of
Excellence in R\&D).

\appendix
{ 
\section{The characteristics of nutation resonance in ferromagnet}
\label{appendixa}
First, we investigate the characteristics of precession and nutation resonance modes in ferromagnet.
Following  Ref.~\cite{Mondal2020nutation}, we write the equation for inverse susceptibility as in Eq.\ (B6) of Ref.\ \cite{Mondal2020nutation}
\begin{align}
\begin{pmatrix}
h_x \\
h_y
\end{pmatrix} & = \frac{1}{\gamma M_0} \begin{pmatrix} \Omega_0 + \alpha\dfrac{\partial }{\partial t} + \eta \dfrac{\partial^2 }{\partial t^2} & - \dfrac{\partial }{\partial t}\\
\dfrac{\partial }{\partial t} &  \Omega_0 + \alpha\dfrac{\partial }{\partial t} + \eta \dfrac{\partial^2 }{\partial t^2}
\end{pmatrix}\begin{pmatrix}
m_x \\
m_y
\end{pmatrix}
\label{Eq1}
\end{align}
We replace $m_x = m_y \propto e^{{ i} \omega t}$ such that $\frac{\partial}{\partial t} \rightarrow { i} \omega$ and $\frac{\partial^2}{\partial t^2} \rightarrow - \omega^2$. Therefore, Eq. (\ref{Eq1}) can be recast as 
\begin{align}
\begin{pmatrix}
h_x \\
h_y
\end{pmatrix} & = \frac{1}{\gamma M_0} \begin{pmatrix} \Omega_0 + { i}\alpha\omega - \eta\omega^2  & - { i} \omega\\
{ i}\omega &  \Omega_0 + { i}\alpha\omega - \eta \omega^2
\end{pmatrix}\begin{pmatrix}
m_x \\
m_y
\end{pmatrix}
\label{Eq2}
\end{align}
Now we employ $h_x = h_y = 0$ and $\alpha = 0$ in order to examine the nature of the eigenmodes, finding a relation between $m_x$ and $m_y$. We obtain two following equations:
\begin{align}\label{Eq3}
    \left(\Omega_0  - \eta\omega^2\right) m_x - { i}\omega m_y & = 0\\
    { i}\omega m_x + \left(\Omega_0 - \eta\omega^2\right) m_y & = 0\label{Eq4}
\end{align}
The Eqs. (\ref{Eq3}) and (\ref{Eq4}) can be written in  concise forms as
\begin{align}
\label{Eq5}
    m_x & = \frac{{ i} \omega }{\Omega_0 - \eta\omega^2} m_y\\ m_y &= -\frac{{ i} \omega}{\Omega_0 - \eta \omega^2}m_x
    \label{Eq6}
\end{align}
The precession and nutation resonance frequencies that were obtained previously in Ref.\ \cite{Mondal2020nutation} are:
\begin{align}
\label{Eq7}
    \omega_{\rm FMR} & = \frac{\sqrt{1 + 4\eta \Omega_0} - 1}{2\eta} \approx \Omega_0 - \eta\Omega_0^2\\
    \omega_{\rm FMNR} & = -\frac{\sqrt{1 + 4\eta \Omega_0} + 1}{2\eta} \approx -\frac{1}{\eta} -\Omega_0 \left(1-  \eta \Omega_0\right)
    \label{Eq8}
\end{align}
\begin{figure}[tbh!]
    \centering
    \includegraphics[scale = 0.32]{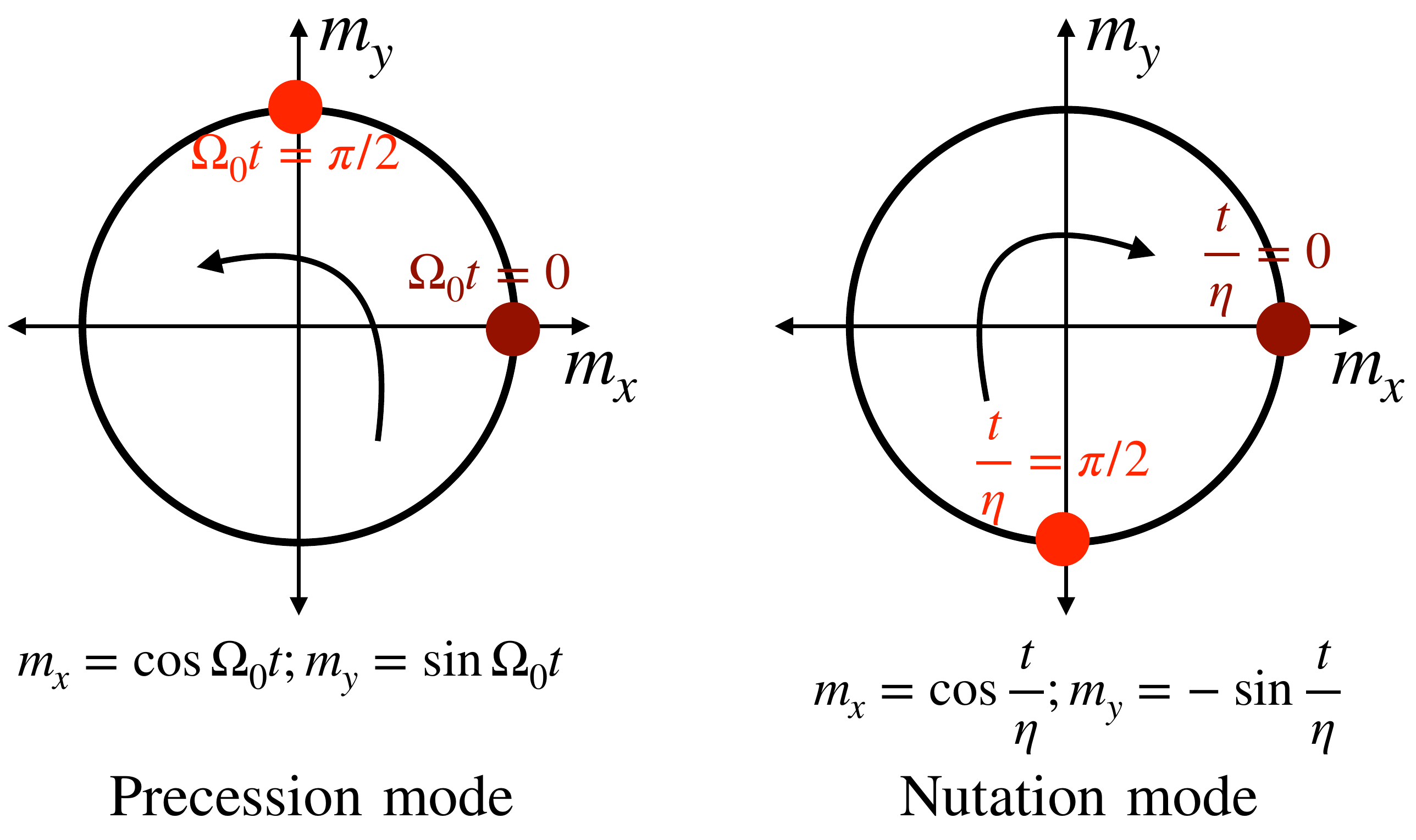}
    \caption{ The precession and nutation mode in ferromagnet. Notice that the precession and nutation modes have opposite sense of rotation.}
    \label{Precession_Nutation_modes}
\end{figure}

We replace the leading order precession resonance term i.e. $\omega_{\rm FMR} \approx \Omega_0$ in Eqs.\ (\ref{Eq5}) and (\ref{Eq6}). Therefore, at the precession resonance we find the relation between $m_x$ and $m_y$ as follows
\begin{align}
    m_x & = \frac{{ i} \Omega_0}{\Omega_0 }m_y = { i} m_y\\ m_y & = -\frac{{ i}\Omega_0}{\Omega_0}m_x = -{ i}m_x
\end{align}
However, the obtained $m_x$ and $m_y$ are complex quantities. In order to obtain the real and time-dependent parts at the precession resonance, we compute:
\begin{align}
    m_x = {\tt Re}\left[m_xe^{i\omega t}\right] = {\tt Re}\left[m_xe^{i\Omega_0 t}\right] = m_x \cos \Omega_0 t\\
    m_y = {\tt Re}\left[m_ye^{i\omega t}\right] = {\tt Re}\left[-im_xe^{i\Omega_0 t}\right] = m_x \sin \Omega_0 t
\end{align}
At the nutation resonance, however, we replace the leading order frequency term i.e., $\omega_{\rm FMNR} \approx - 1/\eta$ and find
\begin{align}
    m_x & = \frac{{ i} (-1/\eta)}{\Omega_0 - 1/\eta} m_y = -\frac{{ i}}{\eta\Omega_0 -1}m_y = im_y\\ 
    m_y & = -\frac{{ i}(-1/\eta)}{\Omega_0 - 1/\eta} m_x = \frac{ i}{\eta \Omega_0 - 1}m_x = -im_x
\end{align}
We again calculate the real and time-dependent parts at the nutation resonance as
\begin{align}
    m_x = {\tt Re}\left[m_xe^{i\omega t}\right] = {\tt Re}\left[m_xe^{-i \dfrac{t}{\eta}}\right] = m_x \cos  \frac{t}{\eta}\\
    m_y = {\tt Re}\left[m_ye^{i\omega t}\right] = {\tt Re}\left[-im_xe^{-i\dfrac{t}{\eta} }\right] = -m_x \sin \frac{ t}{\eta}
\end{align}
Such characteristics of precession and nutation resonance have been shown in Fig.\ \ref{Precession_Nutation_modes}.

}

\bibliographystyle{apsrev4-1}


%

\end{document}